# Measuring the Impact of Introductory Physics Labs on Learning

Carl Wieman[1, 2] and N.G. Holmes[1]

[1]Stanford University, Department of Physics,
382 Via Pueblo Mall, Stanford, CA, 94305
[2]Stanford University, Graduate School of Education,
485 Lasuen Mall, Stanford, CA, 94305

**Abstract.** Our recent study showed that two lab courses, whose goals were exclusively to reinforce material developed in the lecture courses, do not have any impact on exam performance at the 1% level. In this study, we replicated this analysis with a modified version of one of these lab courses whose goals also included modeling, designing experiments, and analyzing and visualizing data. This modified course used the same sets of apparatus as the previous version, but changed the pre-lab and in-lab activities to focus on developing and testing models with data. The study evaluated the impact of these additional goals and activities. We found that they did not affect students' performance on the final exam.
**PACS:** 01.40.G-, 01.40.-d, 01.50.Qb, 01.55.+b

## I. INTRODUCTION

The American Association of Physics Teacher [1] recently released a report of recommendations for learning goals for the undergraduate physics lab curriculum. This document focuses on six areas: modeling, designing experiments, developing technical and practical skills; constructing knowledge, analyzing and visualizing data, and communicating physics. In practice, actual lab courses focus on a wide variety of goals, including developing conceptual understanding of physics content. Seldom do they consider which goals can be uniquely and optimally achieved in labs. Also, there has been little work examining what students are actually learning in introductory labs. In a recent study, we examined the effectiveness of intro labs at meeting the traditional goal of reinforcing physics concepts and content presented in lecture [2]. We found that a lab course with explicit learning goals focused on reinforcing the lecture content, and coordinated with the lecture course, did not have any impact on exam performance with an uncertainty at the 1% level. This was true for two different calculus-based introductory physics courses, a mechanics course and an electricity and magnetism course. This evaluation supports the ongoing discussion in the lab community surrounding the goals and effectiveness of the introductory physics labs.

We recognize that there are many physics courses where the hands-on laboratory portion of the course is embedded within the lecture portion (such as the studio or workshop physics [3-4], SCALE-UP [5-6], or ISLE [7] approaches). In these situations, there are clear improvements in students' learning of the content material over traditional lecture format, but given the integrated format of the courses, it is unclear how or whether the hands-on activities contribute to that learning. In addition, many large universities do not have the resources to support these integrated courses. It is important, therefore, to evaluate how or whether stand-alone lab courses contribute to student learning.

Given the results of the initial study [2], we redesigned the lab portion of the electricity and magnetism course in the subsequent year with a new set of learning goals and restructured pre-lab and in-lab activities. The redesigned lab course used the same physical apparatus as the previous version of the course, so both courses involved investigating the same underlying physics concepts. Our research question for this paper was to evaluate whether shifting the focus of the lab goals and activities affected students' performance in the content course, as measured by the final exam.

## II. MODIFIED LAB COURSE

The goals and activities of the lab course focused on analyzing and interpreting measurements and data, connecting measurement and data with models, and evaluating and refining models based on data and assumptions about the system. The redesign was inspired by an introductory lab framework that engages students in meaningful reflection on their data and results, and using that reflection to iterate to improve their results or knowledge of a system [8-10]. Through this iterative experimentation process, students were seen to engage more deeply with the physical and measurement models involved in experimentation [11], even leading them to reexamine the assumptions about the system and refine physical models seen in class, based on their measurements.

To redesign the course, the existing lab apparatus were evaluated for measureable model limitations, including what level of precision and analysis was necessary for students to observe such limitations. This apparatus and original experiments were quite conventional, covering Coulomb's law, Faraday's law, simple circuits, etc. and are fully listed in [2]. Only minor adjustments to the order of the experiments were needed to present the experimentation, modeling, and analysis goals in a productive sequence. For the first half of the course, pre-lab activities focused on introducing data analysis concepts and tools. These had previously not been part of the activities associated with this sequence of labs. The in-lab activities focused on giving students deliberate practice with those tools and concepts, exploring them in a physical context. For the second half of the course, the activities focused more explicitly on modeling and testing of models. Pre-lab activities used sequences of questions to direct students through the underlying physical models. For example, a resistor-capacitor experiment used the pre-lab to explore why, physically, a capacitor would discharge according to exponential decay, rather than another model. The in-lab activity, correspondingly, involved using measurements to verify the model, but then to refine measurements to explore the limitations of the model in a physical, real-world context. For example, an experiment exploring the magnetic field of a current loop was oriented so that students would discover the off-axis and fringing effects.

Approximately one third of the students enrolled in the lecture course took the optional but associated lab course. This was similar to the original study and for similar reasons: the requirements of the respective majors. The final exam in the lecture course was used as a performance measure of content learning, and not all items were related to the physics principles discussed in the lab. This structure allows us to compare students who did and did not take the lab on content relevant and not relevant to the lab. We do this by taking the ratio between a student's average score on lab-related questions and their score on non-lab-related questions, as in the original study. If the ratio is greater than one, then the student performed better on the lab-related questions than the non-lab-related questions. If the ratio is less than one, the student performed better on the non-lab-related questions than the lab-related ones. Finally, if the ratio is equal to one, then they performed equally well on both types of questions. We then average those ratios across the students who did (or did not) take the lab course. An average ratio that is greater for the lab students than the non-lab students indicates that taking the lab improved the lab students' learning of those topics covered by the lab. If the average ratio is greater for the non-lab students, that suggests that the labs did damage to students' understanding of the content. If the average ratios are the same that implies that the labs had no added benefit to students' understanding of the content, as was found in the previous study [2]. In [2] we present statistical arguments showing the lack of correlation between responses to the different exam questions, and hence why it is implausible that taking the lab would improve performance on both lab and non-lab related questions.

Based on our previous work, the most plausible hypothesis is that there would be no effect, but there are two plausible alternatives:

1. The shift in lab goals and activities provided a distraction and shift in focus that negatively interfered with learning the lecture material.
2. The shift in lab goals and activities, emphasizing a deeper examination of the models, may have enhanced the students' understanding of the lecture materials.

### III. METHODS

Participants were 443 students in an introductory electricity and magnetism course at a large, elite university who completed the final exam for the lecture course. A subset of those students (n=126) was also enrolled in the lab course. The lecture part of the course involved 3 hours per week of in-class lecture plus 2 hours per week in discussion sections led by graduate teaching assistants (TAs). The lab course involved weekly 2-hour lab sessions with approximately 15 students led by a graduate TA. The pre-lab and in-lab activities were designed by a physics education researcher who also coordinated weekly TA training meetings.

At the beginning of the course, students in both groups completed the Conceptual Survey on Electricity and Magnetism (CSEM) [12]. This survey was used to compare the two groups on entering the course. The survey was completed during the discussion sections.

Student scores on 20 multiple choices questions on the final exam were used to evaluate content learning. Two independent raters coded whether each multiple-choice question was related to or not related to a lab experiment. The descriptions of specific activities in the pre-lab and in-lab activities were used for this evaluation. The two raters performed this coding independently, with one of them not looking in detail at the lab activities, and they agreed on 13/20 items. The remaining items were discussed in the context of a more detailed examination of the student activities in the labs, and the researchers quickly reached consensus on all 20. From this coding, 7 out of 20 questions were coded as being related to the lab content, with 13 questions not related to the lab content.

A ratio was calculated for each student comparing their average score on the lab-related questions with their average score on the non-lab related questions. All assumptions for performing an independent-samples t-test on the ratios were met. Three such assumptions involve the study design and measurements: the dependent variable is continuous; the independent variable consists of two independent groups (a lab group and a non-lab group of students); and there is independence of observations (lab students' performance on the final exam does not effect the non-lab students' performance). The final three assumptions relate to the characteristics of the data: there were no significant outliers as evaluated on a histogram of students' ratios for each group; the student ratios were normally distributed for both groups; and there was homogeneity of variances (see data in the results section).

## IV. RESULTS

On average, students in the course correctly answered 14.91 ± 2.51 ($M \pm SD$) out of the 20 final exam multiple choice questions. Scores were normally distributed (skewness = -0.46). We did basic statistical tests of the reliability and validity of these questions as an assessment instrument. We calculated the Kuder-Richardson Formula 20 (KR-20) for the full set of 20 multiple choice items, which is analogous to Cronbach's alpha, but for dichotomous data. This was found to be KR-20 = 0.63 for the exam, which is an acceptable range for an internally consistent (reliable) test, measuring multiple constructs. Correlations between individual questions and the total score ranged from 0.15 to 0.51, with only two questions having $r < 0.2$. This suggests the test items reasonably discriminate between high and low performing students.

As in Ref. [2], there was a significant difference between the two groups on the 20 multiple choice items on the final exam (Table 1), with the lab students outperforming the non-lab students: $t(227.2) = 3.63$, $p < .001$, 95% confidence interval in the difference is [0.43, 1.47]. As noted above, this was not surprising as the students in each group tended to come from different majors. These differences were also reflected in the difference between the two groups on the CSEM pre-test, with the lab students (M=15.96, SD=5.62) outperforming the non-lab students (M =14.42, SD =5.13): $t(219.5) = 2.60$, $p = .010$, 95% confidence interval of the difference is [0.37, 2.71].

There was no significant difference between the two groups on their ratio scores (average lab-item score over non-lab-item score, see Table 1):, 95% confidence interval of the difference is [-0.03, 0.07]. The difference between the two groups was of similar magnitude (at the 1% level), and negligible, as that in the previous study. It is also interesting to note a decrease in these ratios from the previous study for both groups. That is presumably simply due to the variation in the difficulties of the respective exam questions.

These results demonstrate that the students who take the lab come in and leave the course with higher ability in the course content. It is clear, however, that taking the lab course does not interact with this gap.

**TABLE 1.** Student scores on elements of the final exam by group.

|  | # items | Non-lab students (n=317) | Lab students (n=126) |
|---|---|---|---|
| Overall score | 20 | M=73%, SD=12% | M=78%, SD=13% |
| Lab related questions | 7 | M=72%, SD=18% | M=78%, SD=19% |
| Non-lab related questions | 13 | M=74%. SD=13% | M=78%, SD=13% |
| Average Ratios |  | M=0.99, SD=0.26 | M=1.01, SD=0.25 |
| Difference between ratios |  | 0.01±0.05; $t(237.38)=0.73$, $p=.464$ | |

## V. DISCUSSION

In this study, we replicated the methodologies of a recent study that evaluated student learning of course content (as measured by the final exam) in a lab course that focused on reinforcing the course content. The previous study found that the lab provided no added value. In the subsequent year, the lab course was redesigned to incorporate learning goals and activities focused on modeling and experimentation. This study found that the lab, with a greater focus on measurement, experimentation, and modeling, again, offered no added value for learning the course content. The new structure and goals did not help or hinder their learning of this material.

The redesigned lab course, however, offered additional learning experiences, covering additional learning goals. This is in contrast to the previous iteration of the lab course, whether the goals were exclusively targeting course content. We argue, therefore, that this redesign of the lab course did potentially offer added value that was complementary to the lecture course content. We use this result as evidence that refocusing lab courses towards teaching modeling and experimentation does not necessarily come at the cost of learning physics content when there is an associated lecture course.

A limitation of this study, of course, is that we have not evaluated learning in the measurement, experimentation, and modeling aspects of the course. Since there are no common measures or assessments of learning in these areas, this evaluation would require a more elaborate investigation. There is evidence elsewhere that the lab framework used (focusing on reflection and iteration leading to evaluating and refining models) leads to significant changes in students' scientific behaviors [8-10], as well as their motivation, epistemological frames, and identity [10]. Future work will aim to evaluate whether adapting this framework for these experiments and this population produces similar outcomes.

We are also limited by our measurement of learning of the course content, since the final exam was not evaluated for validity or reliability a priori. The measures included in the results suggest that the exam had sufficient discrimination between high and low performing students and was measuring more than a single construct. Unfortunately, the CSEM was not administered at the end of the course, so we cannot use that as an independent measure. The full set of multiple choice items on the final exam were, however, able to discern the same difference between the two groups as the CSEM pre-test. This suggests that there is some reliability that the final exam is measuring electricity and magnetism concepts.

Based on the evidence thus far, however, this paper, in combination with the previous paper [2], demonstrate that a lab course that is distinct from the lecture part of the course does not provide added value to learning the course material. This was true whether the goals were entirely focused on content, or on modeling and experimentation.

## ACKNOWLEDGEMENTS

We are pleased to acknowledge the assistance of Chaya Nanavati and Mark Kasevich in carrying out this work.